\documentclass[preprint,12pt]{elsarticle}

\usepackage{amsmath,amssymb}
\usepackage{graphicx}
\usepackage{booktabs}
\usepackage{bm}
\usepackage[colorlinks=true,allcolors=blue]{hyperref}

\newcommand{\Rev}{\mathcal{R}}
\newcommand{\logR}{\log \Rev}
\newcommand{\rk}{r_{\kappa}}
\newcommand{\Om}{\Omega}
\newcommand{\Nof}{n_t}

\journal{Physica A}

\begin{document}

\begin{frontmatter}

\title{The Ramsey community number as a renormalization-group crossing}

\author{Alexei Vazquez}
\affiliation{organization={Nodes \& Links Ltd}, city={Cambridge}, country={United Kingdom}}

\begin{abstract}
The Ramsey community number $\rk$ is the smallest size at which a network is
better described by communities than by none, under a Bayesian detection rule.
On the diamond hierarchical lattice we show that $\rk$ is an exact
renormalization-group crossing: the block-model sufficient statistics obey a
linear map with eigenvalues $\{bs,b\}$, the degree-corrected evidence density
flows to $\ln K$ at a community fixed point, and $\rk$ is the generation at
which the running evidence clears the detection threshold. Degree correction
advances detection by two generations. We derive $\rk(b,s;q)$ in closed form
for the whole family. Finally, placing on the lattice the Reichardt--Bornholdt
community Hamiltonian---whose ground state is the partition itself---we find an
exact community-ordered phase: below the ferromagnetic critical temperature the
two hubs lock into opposite communities for any resolution $\gamma>0$, a
staggered order that persists as $n\to\infty$. Allowing each nested
sub-community its own label, the optimal partition is a hierarchy of
$q_{\rm opt}\sim\sqrt{n}$ communities, so the number of Potts states that best
describes the network grows with the network. This hierarchy orders thermally
level by level, through a cascade of first-order transitions whose temperatures
fall as $1/\ln q$, so every stable level persists as $n\to\infty$: the emergent
partition is detectable, optimal, and thermodynamically ordered.
\end{abstract}


\end{frontmatter}

\section{Introduction}
\label{sec:intro}

How large must a network grow before its community structure becomes an
objective fact? The question is sharper than it sounds. Communities---groups of
nodes more densely wired among themselves than to the rest---are the most
studied form of mesoscopic organization in networks~\cite{fortunato2010,
fortunato2016}, yet any sufficiently flexible algorithm will happily carve
communities out of structureless graphs. The modern remedy is Bayesian model
comparison~\cite{holland1983,karrer2011,peixoto2017}: fit both a stochastic
block model, which posits groups, and a featureless null, which does not, and
let the evidence ratio
\begin{equation}
  \Rev = \frac{P(\text{partitioned})}{P(\text{unpartitioned})},
  \qquad
  P(\text{split}) = \frac{\Rev}{1+\Rev},
  \label{eq:R}
\end{equation}
decide. A split is \emph{detected} once $P(\text{split})\ge q$ at a prescribed
certainty $q$. In sparse random graphs this competition famously produces a
genuine phase transition in detectability~\cite{decelle2011}. For a
deterministic family of networks growing by local rules, the same rule defines
the \emph{Ramsey community number}~\cite{d7kf-s1cf},
\begin{equation}
  \rk(q) \;=\; \min\{\, n : P(\text{split})\ge q \,\},
  \label{eq:rk}
\end{equation}
the minimal size at which communities become the better description---in the
spirit of Ramsey theory, where a structure is forced to appear once a system is
large enough. A plain ring never breaks ($\rk=\infty$), while a ring with
second-neighbor bonds breaks at $\rk\approx35$~\cite{vazquez2026ringwantsbroken}. On the
scale-free pseudofractal web of Dorogovtsev, Goltsev, and
Mendes~\cite{dgm2002}, degree heterogeneity confounds the question: a plain
block model splits the graph merely to absorb its hubs, and only a
degree-corrected comparison~\cite{karrer2011} isolates genuine communities,
dropping $\rk$ from $1095$ to $42$~\cite{vazquez2026communitystructurepseudofractalweb}. In all cases reported
so far, however, the transition and its asymptotic slope $\logR\sim\sigma n$
were extracted numerically from closed-form evidences: exact inputs, but a
fitted output.

In this work we show that on hierarchical lattices the Ramsey community
number is not a fitted number but an exact renormalization-group (RG)
statement. We work on the diamond hierarchical lattice of Migdal and
Kadanoff~\cite{migdal1976,kadanoff1976}, the classic arena where real-space
decimation is exact and spin models are solved by iterating a
recursion~\cite{berker1979,griffiths1982}. Our core results are threefold. First,
the block-pair sufficient statistics of the natural self-similar
bipartition---edge counts within and between blocks, and block degree
sums---transform under one generation of growth by an exact
\emph{linear} map whose spectrum consists of a volume eigenvalue $bs$ and a
seam eigenvalue $b$, where $b$ is the branching number and $s$ the series
length of the elementary cell. Second, the Bayesian evidence, a fixed nonlinear
function of these statistics, is carried by this flow to a \emph{community
fixed point} at which the evidence density per edge equals $\ln K$, with $K$
the number of communities---a lattice-independent value that doubles as the
fixed-point free-energy density of the detection problem. Third, $\rk$ is
exactly the generation at which the running, extensive evidence first exceeds
the detection threshold, and it admits a closed form proved for the whole
family, with an explicit finite-size correction (Sec.~\ref{sec:general}). Every statistic,
recursion, and evidence below is exact, and each is verified against direct
construction of the lattice.

These detection results have a thermodynamic counterpart on the same lattice.
Because the linear map is literally a real-space RG recursion
(Sec.~\ref{sec:critical}), we can ask whether the community partition is not
only detectable but thermodynamically ordered. Placing the
Reichardt--Bornholdt community Hamiltonian~\cite{reichardt2004}---whose ground
state is the partition itself---on the diamond and solving it by the same
decimation, we find an exact community-ordered phase: below the ferromagnetic
critical temperature the two hubs lock into opposite communities for any
resolution $\gamma>0$, a staggered order that survives as $n\to\infty$
(Sec.~\ref{sec:rb}). Giving each nested sub-community its own label, the
modularity-optimal description is not two communities but a hierarchy of
$q_{\rm opt}\sim\sqrt{n}$ of them, growing with the lattice
(Sec.~\ref{sec:hier}); and this hierarchy orders thermally, level by level,
through a cascade of first-order Potts transitions whose temperatures fall as
$1/\ln q$ (Sec.~\ref{sec:pottsT}). The emergent partition is thus detectable,
optimal, and thermodynamically ordered.

\section{The diamond lattice and the bundle partition}
\label{sec:model}

\begin{figure*}[tb]
  \centering
  \includegraphics[width=\textwidth]{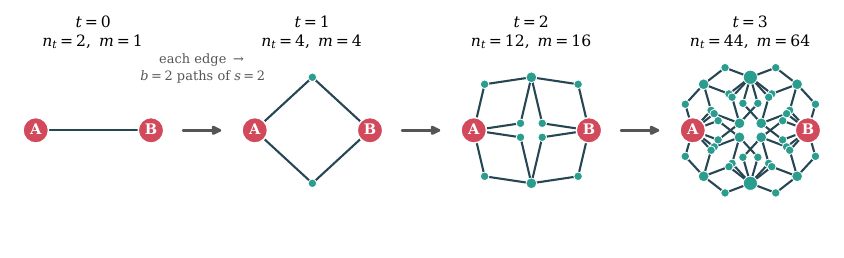}
  \caption{Construction of the diamond hierarchical lattice for $b=s=2$ by
  edge replacement. Starting from a single bond between the poles $A,B$
  ($t=0$), each generation replaces every edge by $b=2$ parallel paths of
  $s=2$ edges, adding one interior vertex per path; iterating gives $m=4^t$
  edges and $\Nof=(2\cdot4^t+4)/3$ nodes [Eq.~\eqref{eq:sizes}]. Node area
  $\propto$ degree, so the two poles emerge as the permanent degree-$2^t$
  hubs.}
  \label{fig:construction}
\end{figure*}

The $(b,s)$ diamond lattice is grown by edge replacement. Starting from a
single bond between two poles $A$ and $B$, each generation replaces every edge
by $b$ parallel paths of $s$ edges; the $s\!-\!1$ interior vertices of every
path are new (Fig.~\ref{fig:construction}). We present the standard diamond
$b=s=2$, which is also the
$(2,2)$-flower of the recursive scale-free nets~\cite{rozenfeld2007}, and note
the general spectrum where it is established. After $t$ generations the lattice
has
\begin{equation}
  m = (bs)^t = 4^t \ \text{edges}, \qquad
  \Nof = \tfrac{1}{3}\left(2\cdot 4^t + 4\right)\ \text{nodes}.
  \label{eq:sizes}
\end{equation}
A vertex born at generation $k$ has degree $2\cdot2^{\,t-k}$, and the poles are
permanent hubs of degree $b^t=2^t$: the lattice is strongly degree heterogeneous,
which is what makes the plain versus degree-corrected comparison nontrivial.

\begin{figure}[tb]
  \centering
  \includegraphics[width=0.80\columnwidth]{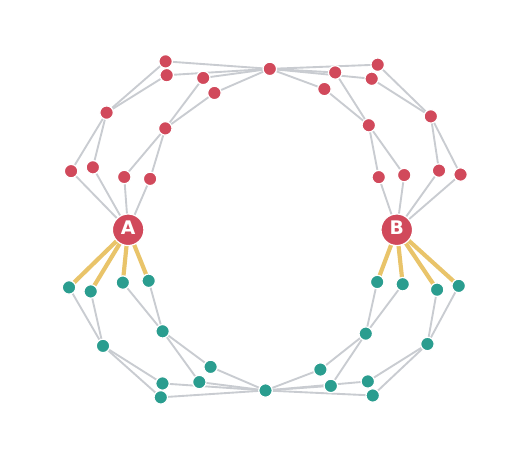}
  \caption{Diamond lattice $G_3$ ($\Nof=44$) with the self-similar bundle cut.
  Red: bundle~0 together with the poles $A,B$; teal: bundle~1; gold: the
  $e_{12}=2^t=8$ seam edges, every one incident on a pole. Node area $\propto$ degree, exposing the two degree-$2^t$ hubs.}
  \label{fig:lattice}
\end{figure}

The self-similar community structure is the pair of parallel bundles between
the poles. Label the two bundles created at $t=1$ and let every later vertex
and edge inherit the label of the edge it replaces. The bipartition we test is
block~$1=$ bundle~$0\cup\{A,B\}$, block~$2=$ bundle~$1$
(Fig.~\ref{fig:lattice}), the analogue of the recursive branch cut of the
pseudofractal web~\cite{vazquez2026communitystructurepseudofractalweb}. Because the bundles meet only at the
poles, block~1 is internally complete and every cross edge is an edge of
bundle~1 incident on a pole. Elementary counting then gives closed forms for
the block-pair statistics---$e_{rs}$ edges between blocks $r$ and $s$, and
block degree sums $\kappa_r$ with $\kappa_1+\kappa_2=2m$:
\begin{align}
  e_{11} &= \tfrac{4^t}{2}, &
  e_{22} &= \tfrac{4^t}{2}-2^t, &
  e_{12} &= 2^t, \notag\\
  \kappa_1 &= 4^t+2^t, &
  \kappa_2 &= 4^t-2^t, &&
  \label{eq:counts}
\end{align}
verified against direct construction through $t=7$ (Table~\ref{tab:counts}).
The seam is subextensive, $e_{12}=2^t=\sqrt{m}$, so the fraction of edges
crossing the cut decays geometrically, $e_{12}/m=(1/s)^t\to0$: the two halves
become ever more cleanly separated as the lattice grows.

\begin{table}[tb]
  \centering
  \caption{Closed-form block-pair statistics of the bundle cut,
  Eq.~\eqref{eq:counts}, verified against direct construction through $t=7$.}
  \label{tab:counts}
  \begin{tabular}{lcccccc}
    \toprule
    $t$ & $\Nof$ & $e_{11}$ & $e_{22}$ & $e_{12}$ & $\kappa_1$ & $\kappa_2$\\
    \midrule
    1 & 4    & 2    & 0    & 2  & 6    & 2   \\
    2 & 12   & 8    & 4    & 4  & 20   & 12  \\
    3 & 44   & 32   & 24   & 8  & 72   & 56  \\
    4 & 172  & 128  & 112  & 16 & 272  & 240 \\
    5 & 684  & 512  & 480  & 32 & 1056 & 992 \\
    \bottomrule
  \end{tabular}
\end{table}

\section{The renormalization-group map}
\label{sec:rgmap}

The central structural observation is that one generation of growth acts on the
sufficient statistics $\bm v=(e_{11},e_{22},e_{12},\kappa_1,\kappa_2)^{\top}$
\emph{exactly linearly}, $\bm v(t{+}1)=M\bm v(t)$, with
\begin{equation}
  M=\begin{pmatrix}
     4&0&0&0&0\\ 0&4&2&0&0\\ 0&0&2&0&0\\ 0&0&-2&4&0\\ 0&0&2&0&4
    \end{pmatrix},
  \quad
  \begin{aligned}
   \lambda_{\rm vol}&=bs=4,\\
   \lambda_{\rm seam}&=b=2 .
  \end{aligned}
  \label{eq:M}
\end{equation}
Read backwards, $M^{-1}$ is the Migdal--Kadanoff decimation that strips the
youngest generation; read forwards, $M$ is the growth generator of the
detection problem. Its entries have a direct physical reading: an edge inside
either block is replaced by four edges whose new interior vertices inherit its
bundle label, so within-block edges quadruple. A seam edge joins a pole
(block~1) to a bundle-1 vertex (block~2), and its new vertices carry the
bundle-1 label into block~2: of its four children, the two at the pole remain
seam edges and the two at the old block-2 endpoint are absorbed into block~2.
Hence $e_{12}'=2e_{12}$ with the displaced pair reappearing as $+2e_{12}$ in
$e_{22}'$; the degree rows follow from $\kappa_r=2e_{rr}+e_{12}$.
The spectrum consists of a fourfold \emph{volume} eigenvalue
$\lambda_{\rm vol}=bs$, which reproduces the growth of the edge count and the
degree sums, and a single \emph{seam} eigenvalue $\lambda_{\rm seam}=b$, which
governs the interface, $e_{12}=b^t$. In the symmetric cell $b=s$ the two roles
cannot be told apart; the general derivation of Sec.~\ref{sec:general} shows
the seam obeys $c_{t+1}=b\,c_t$ exactly for all $(b,s)$, confirmed by direct
construction across the family. The spectral gap is thus
\begin{equation}
  \frac{\lambda_{\rm seam}}{\lambda_{\rm vol}} = \frac{1}{s} ,
  \label{eq:gap}
\end{equation}
set by the series length alone. For $s\ge2$ the seam is an irrelevant direction
relative to the bulk and the crossing fraction flows to zero: the cut is a
genuine, ever-strengthening community boundary. The degenerate case $s=1$
(parallel bonds, no interior vertices) has unit gap and never separates---the
diamond analogue of the plain cycle that never breaks~\cite{vazquez2026ringwantsbroken}.
Community detectability on these lattices is, in this precise sense, a relevant
direction of the RG flow, switched on exactly when $s\ge2$; the two control
parameters play distinct roles, $b$ setting the number of communities and $s$
the sharpness of their separation.

\section{Bayesian evidence for the split}
\label{sec:evidence}

We now let the evidence ride this flow. The detection rule is that of the
companion works~\cite{vazquez2026ringwantsbroken,vazquez2026communitystructurepseudofractalweb}; we recall its degree-corrected
construction in enough detail that Eq.~\eqref{eq:dc} can be read off directly.
Following Ref.~\cite{vazquez2026communitystructurepseudofractalweb}, the graph is modeled as a Poisson multigraph
with degree-corrected rates
$\langle A_{ij}\rangle=\theta_i\theta_j\,\omega_{g_ig_j}$, where
$\theta_i=k_i/\sqrt{2m}$ carries the degree of node $i$ ($2m=\sum_i k_i$) and
$g_i$ is its block. The null $\omega_{rs}\equiv1$ is then the configuration
model---random wiring at fixed degrees~\cite{karrer2011}---so that any split it
favors reflects structure beyond the degree sequence. The affinities then have
configuration-model edge expectations $\Om_{rr}=\kappa_r^2/4m$ and
$\Om_{rs}=\kappa_r\kappa_s/2m$ ($r\neq s$), with $\kappa_r$ the degree sum of
block $r$ and $\sum_{r\le s}\Om_{rs}=m$. The Poisson likelihood factorizes
over block pairs, $P(A\mid\omega,g)=C(A,\bm k)\prod_{r\le s}
\omega_{rs}^{\,e_{rs}}e^{-\omega_{rs}\Om_{rs}}$, and its prefactor $C(A,\bm k)$
depends only on the fixed degrees and adjacency, not on the partition; it is
this factor---the degree-only and combinatorial part---that cancels in any
evidence ratio. Placing a symmetric conjugate prior
$\omega_{rs}\sim\Gamma(\alpha,\alpha)$ (mean one, centered on the configuration
model) and integrating each affinity out leaves the marginal evidence of one
block pair,
\begin{equation}
  Z(e,\Om)=\frac{\alpha^\alpha}{\Gamma(\alpha)}\int_0^\infty\!
    \omega^{\,e+\alpha-1}e^{-(\Om+\alpha)\omega}\,d\omega
    =\frac{\alpha^\alpha}{\Gamma(\alpha)}\,
     \frac{\Gamma(e{+}\alpha)}{(\Om{+}\alpha)^{e+\alpha}} .
  \label{eq:Zdc}
\end{equation}
The evidence for the split is the product of these factors over the occupied
block pairs, divided by the single-block null---one pair comprising the whole
graph, with $e\!\to\!m$ and $\Om\!\to\!m$. Taking logarithms,
$\logR_{\rm dc}=\sum_{r\le s}\ln Z(e_{rs},\Om_{rs})-\ln Z(m,m)$, and inserting
Eq.~\eqref{eq:Zdc},
\begin{align}
  \logR_{\rm dc}
  &= \sum_{r\le s}\Big[\ln\Gamma(e_{rs}{+}\alpha)
      -(e_{rs}{+}\alpha)\ln(\Om_{rs}{+}\alpha)
      +\alpha\ln\alpha-\ln\Gamma(\alpha)\Big] \notag\\
  &\quad -\Big[\ln\Gamma(m{+}\alpha)-(m{+}\alpha)\ln(m{+}\alpha)
      +\alpha\ln\alpha-\ln\Gamma(\alpha)\Big],
  \label{eq:dc}
\end{align}
where the second bracket is the null term $\ln Z(m,m)$---the whole graph as a
single block, so that $e\!\to\!m$ and $\Om\!\to\!m$, with no cross term. The
$\alpha\ln\alpha-\ln\Gamma(\alpha)$ pieces do not fully cancel: the split sum
carries three of them (one per block pair) against the null's one, leaving a net
Occam factor $[\alpha^\alpha/\Gamma(\alpha)]^{2}$ for the two extra
affinities. For contrast we also evaluate the plain Bernoulli
block model with symmetric $B(\alpha,\alpha)$ priors, including the label-prior
factor $B(n_1{+}\alpha,n_2{+}\alpha)$ whose omission spuriously favors any
split~\cite{vazquez2026communitystructurepseudofractalweb}. Inserting the closed forms~\eqref{eq:counts} makes
both evidences explicit functions of $t$; all evaluations use
80-digit arithmetic and match the direct construction exactly
(Table~\ref{tab:evidence}).

\begin{table}[tb]
  \centering
  \caption{Log evidence for the bundle cut at $\alpha=1$ (slopes are
  $\alpha$-independent). The density $\logR_{\rm dc}/m$ converges to
  $\ln K=\ln2=0.693$. The degree-corrected split is favored from $t=2$, the
  plain split only from $t=5$.}
  \label{tab:evidence}
  \begin{tabular}{lrrrr}
    \toprule
    $t$ & $\Nof$ & $\logR_{\rm dc}$ & $\logR_{\rm pl}$ & $\logR_{\rm dc}/m$\\
    \midrule
    2  & 12     & $0.03$    & $-5.60$   & $0.002$\\
    3  & 44     & $16.32$   & $-11.77$  & $0.255$\\
    4  & 172    & $111.86$  & $-6.04$   & $0.437$\\
    5  & 684    & $559.88$  & $85.55$   & $0.547$\\
    6  & 2732   & $2500.2$  & $604.5$   & $0.610$\\
    7  & 10924  & $10597.0$ & $3021.1$  & $0.647$\\
    8  & 43692  & $43738.3$ & $13447.4$ & $0.667$\\
    10 & 699052 & $718680$  & $234124$  & $0.685$\\
    \bottomrule
  \end{tabular}
\end{table}

\begin{figure*}[tb]
  \centering
  \includegraphics[width=1\textwidth]{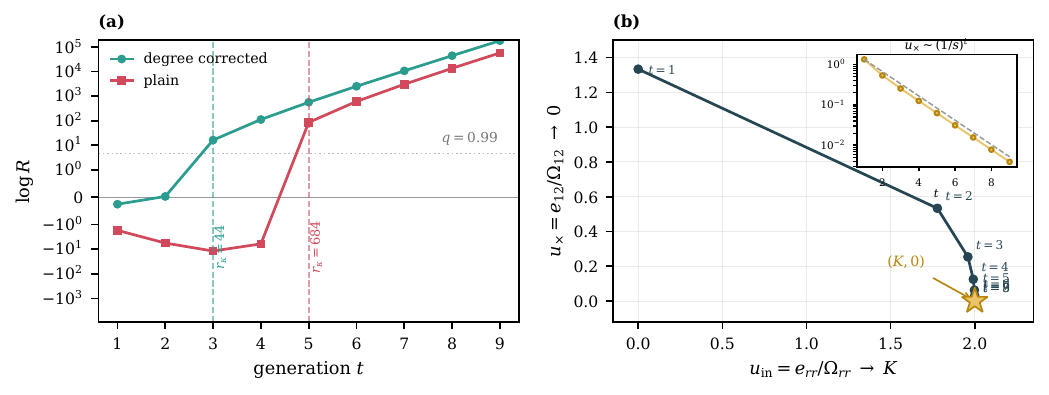}
  \caption{(a) Log evidence of the bundle split versus generation (symlog
  scale). The degree-corrected model clears the detection threshold
  $\ln[q/(1{-}q)]$ (dotted, $q=0.99$) at $t=3$, giving $\rk=n_3=44$; the plain
  model only at $t=5$, $\rk=n_5=684$. The lattice exists only at the discrete
  sizes $\Nof$, so the crossing---and hence $\rk$---is a generation index.
  (b) Renormalization-group flow of the block densities
  Eq.~\eqref{eq:densities} to the community fixed point
  $(u_{\rm in},u_{\times})=(K,0)=(2,0)$; the seam density decays geometrically
  as $(\lambda_{\rm seam}/\lambda_{\rm vol})^t=(1/s)^t$ (inset).}
  \label{fig:data}
\end{figure*}

\section{The fixed-point evidence density}
\label{sec:fixedpoint}

Because $\bm v$ flows under $M$ with dominant rate $bs$ and $\logR_{\rm dc}$ is
a fixed function of $\bm v$, the evidence is carried by the flow, and its
leading behavior is transparent. For large counts,
$\ln\Gamma(e{+}\alpha)-(e{+}\alpha)\ln(\Om{+}\alpha)=e\ln(e/\Om)-e+O(\ln e)$,
and the $-e$ terms cancel against the null since $\sum_{r\le s}e_{rs}=m$. What
survives is the degree-corrected surprise
$\logR_{\rm dc}=\sum_{r\le s}e_{rs}\ln(e_{rs}/\Om_{rs})+O(\ln m)$---the
Kullback--Leibler divergence between the observed and configuration-expected
edge placement, the quantity a degree-corrected modularity
estimates~\cite{newman2006}. Its natural variables are the running block
densities
\begin{equation}
  u_{\rm in}=\frac{e_{rr}}{\Om_{rr}}=\frac{2x^2}{(x\pm1)^2},
  \qquad
  u_{\times}=\frac{e_{12}}{\Om_{12}}=\frac{2x}{x^2-1},
  \label{eq:densities}
\end{equation}
with $x=2^t$. Under the flow they converge to the \emph{community fixed point}
\begin{equation}
  (u_{\rm in},\,u_{\times}) \;\longrightarrow\; (K,\,0) = (2,\,0),
\end{equation}
the seam density decaying as
$(\lambda_{\rm seam}/\lambda_{\rm vol})^t=(1/s)^t$
[Fig.~\ref{fig:data}(b)]. At the fixed point each within-block term contributes
$\tfrac{m}{2}\ln K$ and the cross term is subleading, so the evidence density
converges to a pure number,
\begin{equation}
  \boxed{\ \frac{\logR_{\rm dc}}{m}\ \longrightarrow\ \ln K\ }
  \label{eq:lnK}
\end{equation}
(Table~\ref{tab:evidence}, last column). The interpretation is
information-theoretic. For $K$ equal, well-separated communities,
$e_{rr}=m/K$ while the configuration model expects $\Om_{rr}=m/K^2$; hence
$u_{\rm in}=K$ identically, and each edge contributes exactly $\ln K$---the
entropy of resolving which of $K$ communities it belongs to. The value is
therefore lattice independent, and indeed the symmetric three-community cut of
the pseudofractal web obeys $e_{rr}/\Om_{rr}\to3$ with density
$\ln3$~\cite{vazquez2026communitystructurepseudofractalweb}. Equation~\eqref{eq:lnK} is the fixed-point
free-energy density of the detection problem.

The transient of the flow is equally instructive. At $t=1$ block~2 is a single
edgeless vertex, $u_{\rm in}^{(2)}=0$, and the trajectory in
Fig.~\ref{fig:data}(b) starts far from the fixed point; there the surprise is
small and the $O(\ln m)$ remainder of Eq.~\eqref{eq:dc}---the Occam cost of
the extra block affinities---dominates, making $\logR_{\rm dc}$ negative: a
partition is the more complex hypothesis and must earn its keep. Since the
seam mode decays by $1/s$ per generation, the densities lock onto $(K,0)$
within a few generations and the extensive $\ln K$ term takes over. The prior insensitivity follows likewise: $\alpha$ enters only
through the $O(\ln m)$ Occam terms.

\section{The Ramsey community number}
\label{sec:rk}

The Ramsey community number now follows without any fitting. The rule
$P(\text{split})\ge q$ reads $\logR\ge\ln[q/(1{-}q)]$, and since the lattice
exists only at the discrete sizes $\Nof$, $\rk$ is the first generation to
clear the threshold,
\begin{equation}
  \rk(q)=\Nof\big|_{t_\star}, \qquad
  t_\star=\min\Big\{t:\logR_{\rm dc}(t)\ge\ln\tfrac{q}{1-q}\Big\}.
  \label{eq:cross}
\end{equation}
With $\logR_{\rm dc}\simeq(\ln K)(bs)^t$ from Eq.~\eqref{eq:lnK}, the crossing
generation is
\begin{equation}
  t_\star \simeq \Big\lceil
    \log_{bs}\frac{\ln[q/(1-q)]}{\ln K}
  \Big\rceil ,
  \label{eq:tstar}
\end{equation}
up to a finite-size correction near threshold (Table~\ref{tab:evidence}). This
is the precise sense in
which $\rk$ is an RG crossing: it is the generation at which the running,
extensive evidence---driven by the volume eigenvalue once the flow has turned
onto the community fixed point---first exceeds a fixed detection threshold.
Evaluated exactly [Fig.~\ref{fig:data}(a), Table~\ref{tab:rk}], the
degree-corrected model gives $\rk=12$ at $q=0.5$ and $\rk=44$ from $q=0.9$ all
the way to $q=0.999$: a sharp, $\alpha$-independent step, because the evidence
jumps between discrete sizes. The plain model does not favor the split until
$t=5$, so $\rk=684$ across the same range. Degree correction advances
detection by two generations---an order of magnitude in size---for a
transparent reason. The plain model must describe each block with a single
internal density, yet the blocks are far from homogeneous: half of all edges
are incident on the youngest generation while the poles carry degree $2^t$.
Its likelihood is spent explaining the degree distribution rather than the
cut, and the split pays only once the bundles' mutual avoidance outweighs that
cost---the mechanism behind the hub--leaf degree artifact of the pseudofractal
web~\cite{vazquez2026communitystructurepseudofractalweb}. The configuration null removes this confounder at the outset. The asymptotic slopes
are $\alpha$ independent, $\sigma_{\rm dc}=\ln2=0.693$ against
$\sigma_{\rm pl}\approx0.230$, a ratio close to $3$---\emph{not} the factor of
two found for the pseudofractal, which is therefore not universal.

\begin{table}[tb]
  \centering
  \caption{Ramsey community number $\rk(q)=\Nof|_{t_\star}$ for the bundle cut
  ($\alpha=1$). Degree correction advances detection from generation 5 to 3.}
  \label{tab:rk}
  \begin{tabular}{lcc}
    \toprule
    $q$ & degree corrected & plain \\
    \midrule
    $0.5$   & $n_2=12$ & $n_5=684$\\
    $0.9$--$0.999$ & $n_3=44$ & $n_5=684$\\
    \bottomrule
  \end{tabular}
\end{table}

\section{The general \texorpdfstring{$(b,s)$}{(b,s)} family}
\label{sec:general}

The construction generalizes across the $(b,s)$ family, with
$\lambda_{\rm vol}=bs$, $\lambda_{\rm seam}=b$, gap $1/s$, and, for the
symmetric partition into the $b$ bundles, fixed-point density $\ln K=\ln b$.
With $n_t=2+b(s{-}1)[(bs)^t{-}1]/(bs{-}1)$ nodes at generation $t$,
Eq.~\eqref{eq:tstar} at $K=b$ delivers the central object of this work in
closed form,
\begin{align}
  \rk(b,s;q) &= 2+\frac{b(s{-}1)}{bs-1}\Big[(bs)^{t_\star}-1\Big],
  \label{eq:rkbs}\\
  t_\star &= \Big\lceil \log_{bs}\!\Big(\ln\tfrac{q}{1-q}\Big/\ln b\Big)
  \Big\rceil .\notag
\end{align}
Here $b$ enters through the entropy accumulated per edge and $s$
through the volume growth, while degrees, priors, and lattice details drop
out. We now derive this closed form, and the exact linear map and fixed-point
density on which it rests, for the whole family.

\subsection{Bundle lemma}
Label the $b$ paths created at $t=1$ as bundles
$0,\dots,b{-}1$, and let every vertex and edge created later inherit the
bundle label of the edge it replaces. We claim that, at every generation and
for $s\ge2$, each edge either has both endpoints interior to a single bundle
or joins a pole to a bundle vertex; no edge ever joins two distinct bundles.
This holds at $t=1$, where the bundles are internally disjoint paths sharing
only the poles. Inductively, replacing an edge creates $s\!-\!1$ interior
vertices per path, all carrying the parent's bundle label, and every child
edge joins two vertices from the set \{parent endpoints, new vertices\}; since
the parent's endpoints belong to one bundle (or are poles) and the new
vertices inherit its label, no child can bridge two bundles. $\square$

\subsection{Per-bundle statistics and the linear map}
By the lemma, all
block-pair statistics of the $K{=}b$ bundle partition (block $0=$ bundle $0\cup
\{A,B\}$, block $r=$ bundle $r$) are linear in two per-bundle counts: the
pole-incident edges $c_t$ and the interior edges $i_t$ of one bundle. Under
one generation, a pole-incident edge spawns $b$ paths, each contributing
exactly one new pole-incident edge and $s\!-\!1$ interior edges, while an
interior edge spawns $bs$ interior edges:
\begin{equation}
  c_{t+1}=b\,c_t, \qquad i_{t+1}=bs\,i_t+b(s{-}1)\,c_t .
  \label{eq:recbs}
\end{equation}
The map is triangular, so its spectrum is read off the diagonal:
$\lambda_{\rm vol}=bs$ and $\lambda_{\rm seam}=b$, proving
the spectrum of Eq.~\eqref{eq:M} for all $(b,s)$. With the initial condition of a
single path per bundle, $c_1=2$, $i_1=s-2$, the solution is
\begin{equation}
  c_t=2\,b^{\,t-1}, \qquad i_t=b^{\,t-1}\!\left(s^t-2\right)
  \qquad (s\ge2).
  \label{eq:icbs}
\end{equation}
The block statistics follow by the lemma: $e_{00}=i_t+c_t=b^{\,t-1}s^t$,
$e_{rr}=i_t$ and $e_{0r}=c_t$ for $r\ge1$, and $e_{rs}=0$ for $r\neq s\ge1$;
the block degree sums are $\kappa_r=2i_t+c_t=2b^{\,t-1}(s^t{-}1)$ and
$\kappa_0=\kappa_r+b\,c_t=2b^{\,t-1}(s^t{-}1{+}b)$; and since each replaced
edge adds $b(s{-}1)$ vertices, $n_t=2+b(s{-}1)[(bs)^t{-}1]/(bs{-}1)$. All of
these, including $e_{rs}=0$, were checked against direct construction for
$(b,s)$ with $b\le4$, $s\le5$. The $b=s=2$ specialization reproduces
Eqs.~\eqref{eq:counts} and~\eqref{eq:M} (there block $2$ collects the second
bundle, $e_{22}=i_t$, $e_{12}=c_t$).

\subsection{Densities and the fixed point}
Writing $x=s^t$, the configuration
ratios are exact rational functions,
\begin{align}
  \frac{e_{00}}{\Om_{00}} &= \frac{b\,x^2}{(x{+}b{-}1)^2}, &
  \frac{e_{rr}}{\Om_{rr}} &= \frac{b\,x(x{-}2)}{(x{-}1)^2}, \notag\\
  \frac{e_{0r}}{\Om_{0r}} &= \frac{b\,x}{(x{-}1)(x{+}b{-}1)}
  &&\xrightarrow{\ t\to\infty\ } \frac{b}{s^t},
  \label{eq:ratiosbs}
\end{align}
so both within-block densities flow to $u_{\rm in}=b=K$ and the seam density
to zero as $(1/s)^t$, establishing the community fixed point $(K,0)$ for the
whole family.

\subsection{Evidence density and its correction}
Insert
Eq.~\eqref{eq:ratiosbs} into the surprise form of the evidence,
$\logR_{\rm dc}=\sum_{r\le s}e_{rs}\ln(e_{rs}/\Om_{rs})+O(\ln m)$; the empty
pairs $e_{rs}=0$ ($r\neq s\ge1$) contribute only their Occam terms,
$O(b^2\ln m)$. Because every ratio in Eq.~\eqref{eq:ratiosbs} carries the same
prefactor $b$ and $\sum_{r\le s}e_{rs}=m$, the $\ln b$ contributions sum to
$m\ln b$ \emph{exactly}; expanding the residual logarithms in $1/x$ gives
\begin{align}
  \logR_{\rm dc} &= m\ln b\left[\,1-\delta_t\,\right],
  \label{eq:delta}\\
  \delta_t &= \frac{2(b{-}1)}{b\ln b}\,\frac{t\ln s+1}{s^{t}}
  +O\!\Big(\frac{t}{s^{2t}}\Big),\notag
\end{align}
up to the $O(m^{-1}\ln m)$ Occam terms. The dominant deficit is the seam's:
$e_{0r}\ln(e_{0r}/\Om_{0r})\simeq -c_t\,t\ln s$ per pair, i.e., the price of
the $b^t$ crossing edges still present at generation $t$. Numerically the
ratio of the exact deficit to Eq.~\eqref{eq:delta} is $1.027$, $1.002$,
$1.0001$ at $t=6,10,14$ for $(b,s)=(2,2)$, and analogously across the family.

\subsection{Validity of the crossing formula}
Let $L=\ln[q/(1{-}q)]$. By
Eq.~\eqref{eq:delta} the exact crossing generation equals the ceiling formula
of Eq.~\eqref{eq:rkbs} whenever $L\le(1-\delta_{t_\star})\,(\ln b)(bs)^{t_\star}$,
and exceeds it by exactly one generation otherwise: the misfire window has
relative width $\delta_{t_\star}\sim t_\star\ln s/s^{t_\star}$ and vanishes as
$q\to1$. Both regimes occur: for $(b,s)=(2,2)$ the formula is exact at every
decade $10^2\le L\le10^8$, while for $(3,2)$ at $L=10^3$ the evidence at the
predicted generation is $994.1$, just short, and the crossing lands one
generation later---precisely the mechanism quantified by
Eq.~\eqref{eq:delta}.

\section{Connection to critical phenomena}
\label{sec:critical}

The dictionary with critical
phenomena on these lattices~\cite{migdal1976,kadanoff1976,berker1979,
griffiths1982} is direct: the sufficient statistics are the couplings, $M$ the
recursion, the volume and seam eigenvalues the bulk and interfacial rescaling
factors, detectability a relevant direction of the flow for $s\ge2$, and
$\ln K$ the fixed-point free-energy density. This correspondence is not merely
formal: on the same lattice one can place a spin model whose ground state is the
community partition itself, and ask whether community order survives
thermally---which we take up next.

\section{A thermodynamic community-ordered phase}
\label{sec:rb}

The correspondence of Sec.~\ref{sec:critical} is with the \emph{detection} free
energy. We now show that the community partition is also the ground state of a
genuine spin model on the lattice, and that it supports an exact
\emph{thermodynamic} ordered phase that persists as $n\to\infty$. The natural
Hamiltonian is that of Reichardt and Bornholdt~\cite{reichardt2004}, whose
minimization is community detection and which reduces to modularity at unit
resolution,
\begin{equation}
  \mathcal{H} = -\sum_{i\neq j}\big(A_{ij}-\gamma\,p_{ij}\big)\,
                \delta(\sigma_i,\sigma_j),
  \label{eq:rbH}
\end{equation}
with $A_{ij}$ the adjacency matrix, $\sigma_i\in\{1,\dots,q\}$ the community
label, $\gamma$ a resolution parameter, and $p_{ij}=\kappa_i\kappa_j/2m$ the
configuration null. We take the Ising case $q=2$, $\sigma_i=\pm1$. Using
$\delta(\sigma_i,\sigma_j)=\tfrac12(1+\sigma_i\sigma_j)$ and dropping constants,
Eq.~\eqref{eq:rbH} becomes a ferromagnet on the graph edges plus an
infinite-range, degree-weighted antiferromagnetic penalty,
\begin{equation}
  \mathcal{H} = -\!\sum_{\langle ij\rangle}\!\sigma_i\sigma_j
    \;+\;\frac{\gamma}{4m}\,\Big[M_\kappa^2-\!\sum_i\kappa_i^2\Big],
    \qquad M_\kappa=\sum_i \kappa_i\sigma_i .
  \label{eq:rbIsing}
\end{equation}

\begin{figure*}[t]
  \centering
  \includegraphics[width=1\textwidth]{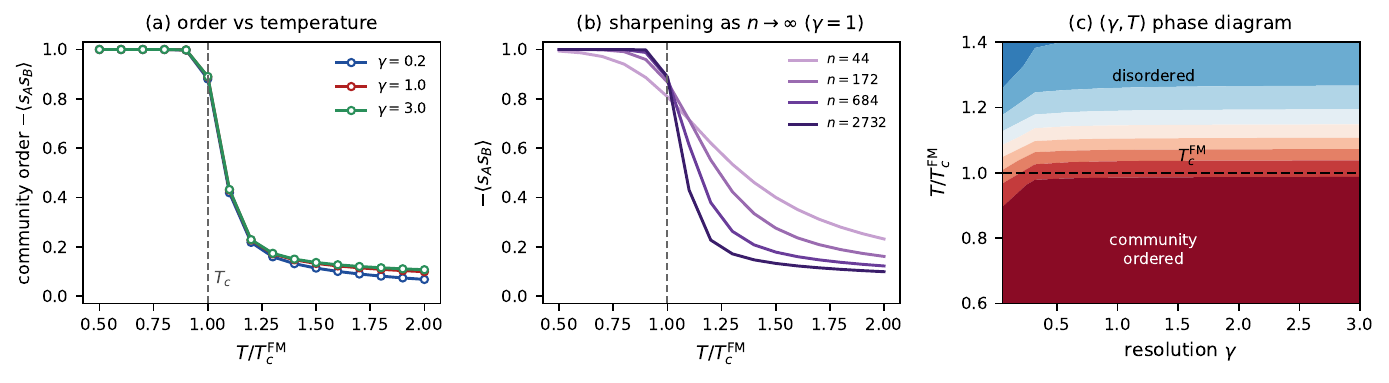}
  \caption{\textbf{Exact community-ordered phase of the $q=2$
  Reichardt--Bornholdt model on the diamond.} Order parameter
  $-\langle\sigma_A\sigma_B\rangle$ (hub anti-alignment: $1$ when the two hubs
  occupy opposite communities). (a)~Versus temperature at $t=6$ ($\Nof=2732$)
  for three resolutions $\gamma$: the community order switches off at the
  ferromagnetic critical temperature $T_c^{\rm FM}$, essentially independent of
  $\gamma$. (b)~As the lattice grows the transition sharpens into a step at
  $T_c^{\rm FM}$; below it the order is persistent, $-\langle\sigma_A\sigma_B
  \rangle\to1$ as $n\to\infty$. (c)~The $(\gamma,T)$ phase diagram: a
  community-ordered region for all $\gamma>0$ below $T_c^{\rm FM}$ (dashed).}
  \label{fig:rb}
\end{figure*}

The first term wants a single community (all aligned); the second penalizes any
net degree-weighted magnetization, rewarding a \emph{balanced bipartition}.
Their competition is exactly the resolution the parameter $\gamma$ controls.

Because the penalty couples the spins only through the single collective
variable $M_\kappa$, the model is exactly solvable: the ferromagnetic part is
local and handled by the same decimation as above, while the value of
$M_\kappa$ is tracked by an auxiliary weight, giving the partition function as a
sum over generation-resolved edge configurations at $80$-digit precision. We
verified the solution against direct enumeration through $t=3$.

The order parameter is the community assignment of the two hubs: the pole
correlation $\langle\sigma_A\sigma_B\rangle$, which is $+1$ when the poles share
a community (the ferromagnet, one community) and $-1$ when they are split into
opposite communities (the top-level cut of Sec.~\ref{sec:model}). The exact
solution gives a sharp transition (Fig.~\ref{fig:rb}). Below the ferromagnetic
critical temperature $T_c^{\rm FM}$ ($k_BT_c^{\rm FM}/J=1.641$) the two hubs lock into
\emph{opposite} communities, $-\langle\sigma_A\sigma_B\rangle\to1$, and this
holds for \emph{any} resolution $\gamma>0$---even infinitesimal $\gamma$ drives
the split, because the degree-weighted penalty acts most strongly on the two
maximal-degree hubs. As the lattice grows the order parameter sharpens into a
step at $T_c^{\rm FM}$ [Fig.~\ref{fig:rb}(b)]: below it the community order is
\emph{persistent}, $-\langle\sigma_A\sigma_B\rangle\to1$ as $n\to\infty$; above
it the order decays to zero. The community-ordering temperature coincides with
$T_c^{\rm FM}$ and is essentially $\gamma$-independent
[Fig.~\ref{fig:rb}(c)]---the resolution selects \emph{which} partition orders,
not \emph{whether} it does.

This is the thermodynamic counterpart of the detection result. A purely
ferromagnetic Ising model on the diamond has no such phase---its ground state is
the trivial single community, and any community-staggered order parameter
averages to zero. The Reichardt--Bornholdt penalty changes the ground state
itself to the balanced partition, and the self-similar geometry then makes that
partition thermodynamically stable in the limit of large size: the emergent
communities are not merely detectable at $\rk$, they order.

\section{Hierarchical community structure and a growing number of states}
\label{sec:hier}

The two-community cut is only the top of a self-similar tree. Because the
diamond is built by recursion, the same argument that favors splitting the
whole lattice favors splitting each bundle again, and the Reichardt--Bornholdt
functional keeps improving as the partition is refined---provided each nested
sub-community is given its \emph{own} label, so that the number of Potts states
grows with the depth of the hierarchy. At depth $d$ the lattice is cut along the
seams of generations $1,\dots,d$ into $q=2^{d}$ blocks, and the model of
Eq.~\eqref{eq:rbH} is evaluated with $q=2^d$ states.

\begin{figure}[tb]
  \centering
  \includegraphics[width=\columnwidth]{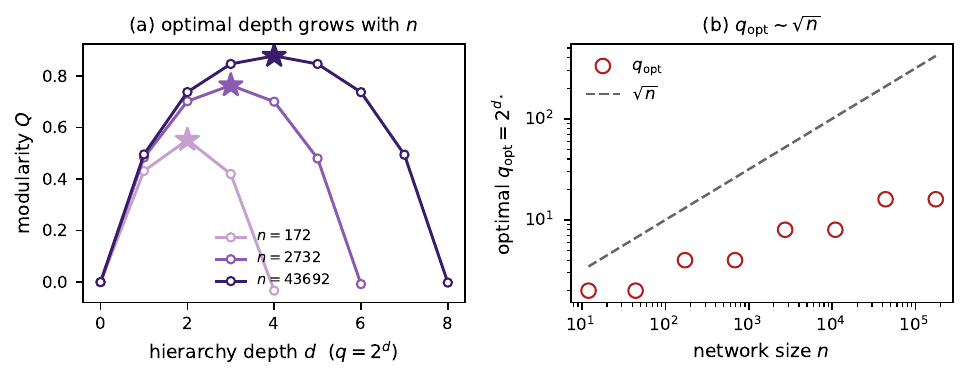}
  \caption{\textbf{Hierarchical community structure with a growing number of
  states.} (a)~Modularity $Q$ of the depth-$d$ nested partition
  ($q=2^d$ communities), Eq.~\eqref{eq:Qd}, versus depth for three sizes; stars
  mark the optimum, which moves to finer resolution as $n$ grows. (b)~The
  optimal number of communities $q_{\rm opt}=2^{d_\star}$ tracks $\sqrt{n}$
  (dashed), Eq.~\eqref{eq:qopt}, over $t\le9$.}
  \label{fig:hier}
\end{figure}

The energetics are exactly solvable. Removing the depth-$d$ seams cuts
\begin{equation}
  c(d) = 2^{\,t+d}\quad(d\ge1),\qquad
  \frac{e_{\rm in}}{m}=1-2^{\,d-t},
  \label{eq:cutd}
\end{equation}
edges---each deeper level doubles the interface, exactly the seam eigenvalue
$\lambda_{\rm seam}=b=2$ acting once per level---while the degree-weighted
penalty of the $2^d$ blocks is $\sum_c\kappa_c^2/(2m)^2\to2^{-d}$ as
$t\to\infty$, dominated by the block that still holds the two hubs. The
modularity of the depth-$d$ partition is therefore, to leading order in large
size,
\begin{equation}
  Q(d) = \frac{e_{\rm in}}{m}-\gamma\,\frac{\sum_c\kappa_c^2}{(2m)^2}
       = 1 - 2^{\,d-t} - \gamma\,2^{-d} + O(2^{-t}).
  \label{eq:Qd}
\end{equation}
The two competing terms are the price of the cut, $2^{d-t}$, growing with
resolution, and the reward for balancing degree, $\gamma\,2^{-d}$, shrinking
with it. Maximizing Eq.~\eqref{eq:Qd} over $d$ gives $2^{2d}=\gamma\,2^{t}$,
i.e.\ an optimal depth and number of communities
\begin{equation}
  d_\star = \tfrac12\big(t+\log_2\gamma\big),
  \qquad
  q_{\rm opt}=2^{d_\star}=\sqrt{\gamma}\;2^{t/2}\sim\sqrt{\gamma\,n},
  \label{eq:qopt}
\end{equation}
with maximal modularity $Q_\star = 1 - 2\sqrt{\gamma}\,2^{-t/2}\to1$. The best
description of the diamond is thus not two communities but a nested hierarchy of
order $\sqrt{n}$ communities of $\sim\sqrt{n}$ nodes each---the geometric
midpoint of its own hierarchy---and the number of Potts states that optimally
describes the network \emph{grows with the network}, as $q_{\rm opt}\sim\sqrt n$
(Fig.~\ref{fig:hier}). We confirm $d_\star=\lfloor t/2\rfloor$ at $\gamma=1$ by
exact construction through $t=9$.

This is the diamond counterpart of the $K_{\rm opt}\sim\sqrt{n}$ hierarchy of
the pseudofractal web~\cite{vazquez2026communitystructurepseudofractalweb}, reached here inside a spin model: the
resolution $\gamma$ sets the depth through Eq.~\eqref{eq:qopt}, each level adds
an interface that costs one factor of the seam eigenvalue, and the single
bundle cut of Sec.~\ref{sec:rb} is recovered as the $d=1$ top of the tree. The
Ramsey community number marks where the first level of this hierarchy becomes
detectable; the deeper levels switch on, one after another, as the lattice
continues to grow.

\section{Finite-temperature order of the hierarchy}
\label{sec:pottsT}

Does the hierarchical community structure survive thermally, as the two-block
order of Sec.~\ref{sec:rb} did? The relevant object is now the $q$-state Potts
model with $q=2^d$ colors, one per block of the depth-$d$ hierarchy. Its
ferromagnetic part---colors align across edges, the intended partition being
the ordered state---is again exactly solvable by decimation: a Potts bond is
described by a same-color weight $A$ and a different-color weight $B$, and one
generation acts by the exact series and parallel laws
\begin{equation}
\begin{aligned}
  A_{\rm ser} &= A_1A_2 + (q{-}1)B_1B_2, \\
  B_{\rm ser} &= A_1B_2 + B_1A_2 + (q{-}2)B_1B_2,
\end{aligned}
\qquad
\begin{aligned}
  A_{\rm par} &= A_1A_2,\\
  B_{\rm par} &= B_1B_2,
\end{aligned}
  \label{eq:pottsRG}
\end{equation}
so the community order parameter---the probability that the two hubs carry
different colors, in excess of chance---follows in closed form at every size.

\begin{figure}[tb]
  \centering
  \includegraphics[width=\columnwidth]{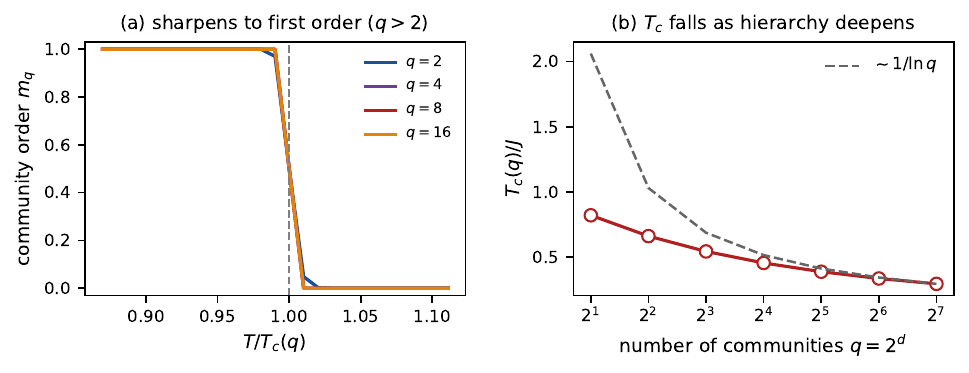}
  \caption{\textbf{Finite-temperature order of the growing-$q$ hierarchy.}
  (a)~Community order parameter $m_q$ (hubs in different communities, in excess
  of chance) versus $T/T_c(q)$ for the $q$-state Potts model, $q=2^d$, at
  $\Nof$ large. The transition is continuous for $q=2$ and sharpens into a
  first-order jump for $q>2$. (b)~The ordering temperature $T_c(q)$ falls as the
  hierarchy deepens, $\sim1/\ln q$ (dashed), so finer communities order at lower
  temperatures.}
  \label{fig:potts}
\end{figure}

The result is a persistent, and increasingly sharp, ordered phase
(Fig.~\ref{fig:potts}). Below a $q$-dependent critical temperature $T_c(q)$ the
hubs occupy different communities and the order is complete as $n\to\infty$;
above it the colors are random. For $q=2$ the transition is continuous---this
is the Ising community order of Sec.~\ref{sec:rb}---but for every $q>2$ it is
\emph{first order}: the order parameter jumps discontinuously at $T_c(q)$, from
near zero to near one across a vanishing temperature window, exactly the
behavior known for the Potts model on hierarchical
lattices~\cite{griffiths1982}. As the hierarchy deepens the transition
temperature falls,
\begin{equation}
  \frac{k_BT_c(q)}{J}\;\sim\;\frac{1}{c\,\ln q},\qquad c\approx0.70,
  \label{eq:Tcq}
\end{equation}
so along the optimal locus $q_{\rm opt}\sim\sqrt{n}$ of Sec.~\ref{sec:hier} the
ordering temperature drifts slowly to zero, $k_BT_c/J\sim1/\ln n$: the deepest
resolvable communities order only at the lowest temperatures, while the coarse
top-level split orders first and remains ordered over the widest range. The
hierarchy thus orders level by level, each finer level switching on at a lower
temperature, but every level that is thermodynamically stable stays ordered as
$n\to\infty$.

The community structure of the diamond therefore has a complete thermodynamic
counterpart. It is detectable above the Ramsey community number
$\rk$ (Sec.~\ref{sec:rk}); its coarsest split is the ground state of the
Reichardt--Bornholdt Hamiltonian and orders below $T_c^{\rm FM}$
(Sec.~\ref{sec:rb}); the optimal partition is a nested hierarchy of
$\sim\sqrt{n}$ communities (Sec.~\ref{sec:hier}); and that hierarchy orders
thermally, level by level, through a cascade of first-order transitions whose
temperatures fall as $1/\ln q$ with the depth. The emergent partition is not a
statistical artifact but a genuine, and persistent, form of order.

\section{Conclusions}
\label{sec:conclusions}

Within this exactly solvable arena the conclusion is complete: the Ramsey
community number is not a fitted slope but a renormalization-group crossing,
the generation at which the extensive evidence---driven by the volume
eigenvalue of the exact map $M$---first clears the detection threshold. On the
same lattice the Reichardt--Bornholdt community Hamiltonian exhibits an exact
thermodynamic community-ordered phase, in which the two hubs occupy opposite
communities below the ferromagnetic critical temperature and the order persists
as $n\to\infty$, so the emergent partition is not only detectable but
thermodynamically stable.

Several directions follow. The nested hierarchy found here, with its
$q_{\rm opt}\sim\sqrt{n}$ optimal number of communities, mirrors the
$K_{\rm opt}\sim\sqrt{n}$ hierarchy of the pseudofractal
web~\cite{vazquez2026communitystructurepseudofractalweb}, suggesting a general law for self-similar graphs that
deserves a proof at the level of the full partition free energy. Apollonian
networks and the $(u,v)$-flowers~\cite{rozenfeld2007} carry the same
construction, mapping $\rk$ across branching numbers. And whether $M$ survives,
exactly or on average, off the deterministic lattice sets how far the RG
reading of $\rk$ extends toward disordered networks, where detectability is
itself a genuine phase transition~\cite{decelle2011}.

\section*{Acknowledgements}
The calculations, computer scripts, and text of this manuscript were generated
with the assistance of Claude Opus 4.8 (Anthropic).

\vspace{4pt}
\bibliographystyle{elsarticle-num}
\bibliography{refs}

@article{dgm2002,
  author = {Dorogovtsev, S. N. and Goltsev, A. V. and Mendes, J. F. F.},
  title = {Pseudofractal scale-free web},
  journal = {Phys. Rev. E},
  volume = {65},
  pages = {066122},
  year = {2002},
  publisher = {American Physical Society},
  doi = {10.1103/PhysRevE.65.066122},
  url = {https://doi.org/10.1103/PhysRevE.65.066122}
}

@article{rozenfeld2007,
  author = {Rozenfeld, Hern\'an D. and Havlin, Shlomo and ben-Avraham, Daniel},
  title = {Fractal and transfractal recursive scale-free nets},
  journal = {New J. Phys.},
  volume = {9},
  pages = {175},
  year = {2007},
  doi = {10.1088/1367-2630/9/6/175},
  url = {https://doi.org/10.1088/1367-2630/9/6/175}
}

@article{holland1983,
  author = {Holland, Paul W. and Laskey, Kathryn Blackmond and Leinhardt, Samuel},
  title = {Stochastic blockmodels: First steps},
  journal = {Social Networks},
  volume = {5},
  pages = {109},
  year = {1983},
  doi = {10.1016/0378-8733(83)90021-7},
  url = {https://doi.org/10.1016/0378-8733(83)90021-7}
}

@article{karrer2011,
  author = {Karrer, Brian and Newman, M. E. J.},
  title = {Stochastic blockmodels and community structure in networks},
  journal = {Phys. Rev. E},
  volume = {83},
  pages = {016107},
  year = {2011},
  publisher = {American Physical Society},
  doi = {10.1103/PhysRevE.83.016107},
  url = {https://doi.org/10.1103/PhysRevE.83.016107}
}

@article{peixoto2017,
  author = {Peixoto, Tiago P.},
  title = {Nonparametric {Bayesian} inference of the microcanonical stochastic block model},
  journal = {Phys. Rev. E},
  volume = {95},
  pages = {012317},
  year = {2017},
  publisher = {American Physical Society},
  doi = {10.1103/PhysRevE.95.012317},
  url = {https://doi.org/10.1103/PhysRevE.95.012317}
}

@article{newman2006,
  author = {Newman, M. E. J.},
  title = {Modularity and community structure in networks},
  journal = {Proc. Natl. Acad. Sci. USA},
  volume = {103},
  pages = {8577},
  year = {2006},
  doi = {10.1073/pnas.0601602103},
  url = {https://doi.org/10.1073/pnas.0601602103}
}

@article{fortunato2010,
  author = {Fortunato, Santo},
  title = {Community detection in graphs},
  journal = {Phys. Rep.},
  volume = {486},
  pages = {75},
  year = {2010},
  doi = {10.1016/j.physrep.2009.11.002},
  url = {https://doi.org/10.1016/j.physrep.2009.11.002}
}

@article{fortunato2016,
  author = {Fortunato, Santo and Hric, Darko},
  title = {Community detection in networks: A user guide},
  journal = {Phys. Rep.},
  volume = {659},
  pages = {1},
  year = {2016},
  doi = {10.1016/j.physrep.2016.09.002},
  url = {https://doi.org/10.1016/j.physrep.2016.09.002}
}

@article{migdal1976,
  author = {Migdal, A. A.},
  title = {Phase transitions in gauge and spin-lattice systems},
  journal = {Sov. Phys. JETP},
  volume = {42},
  pages = {743},
  year = {1976}
}

@article{kadanoff1976,
  author = {Kadanoff, Leo P.},
  title = {Notes on {Migdal's} recursion formulas},
  journal = {Ann. Phys. (N.Y.)},
  volume = {100},
  pages = {359},
  year = {1976},
  doi = {10.1016/0003-4916(76)90066-X},
  url = {https://doi.org/10.1016/0003-4916(76)90066-X}
}

@article{berker1979,
  author = {Berker, A. Nihat and Ostlund, Stellan},
  title = {Renormalisation-group calculations of finite systems: order parameter and specific heat for epitaxial ordering},
  journal = {J. Phys. C: Solid State Phys.},
  volume = {12},
  pages = {4961},
  year = {1979},
  doi = {10.1088/0022-3719/12/22/035},
  url = {https://doi.org/10.1088/0022-3719/12/22/035}
}

@article{griffiths1982,
  author = {Griffiths, Robert B. and Kaufman, Miron},
  title = {Spin systems on hierarchical lattices. {Introduction} and thermodynamic limit},
  journal = {Phys. Rev. B},
  volume = {26},
  pages = {5022},
  year = {1982},
  publisher = {American Physical Society},
  doi = {10.1103/PhysRevB.26.5022},
  url = {https://doi.org/10.1103/PhysRevB.26.5022}
}

@article{decelle2011,
  author = {Decelle, Aurelien and Krzakala, Florent and Moore, Cristopher and Zdeborov\'a, Lenka},
  title = {Inference and phase transitions in the detection of modules in sparse networks},
  journal = {Phys. Rev. Lett.},
  volume = {107},
  pages = {065701},
  year = {2011},
  publisher = {American Physical Society},
  doi = {10.1103/PhysRevLett.107.065701},
  url = {https://doi.org/10.1103/PhysRevLett.107.065701}
}

@article{d7kf-s1cf,
  title = {Emergence of network communities driven by local rules},
  author = {Vazquez, Alexei},
  journal = {Phys. Rev. E},
  volume = {111},
  issue = {6},
  pages = {064314},
  numpages = {9},
  year = {2025},
  month = {Jun},
  publisher = {American Physical Society},
  doi = {10.1103/d7kf-s1cf},
  url = {https://link.aps.org/doi/10.1103/d7kf-s1cf}
}

@misc{vazquez2026ringwantsbroken,
  title = {The ring wants to be broken},
  author = {Vazquez, Alexei},
  year = {2026},
  eprint = {2607.01967},
  archivePrefix = {arXiv},
  primaryClass = {physics.soc-ph},
  url = {https://arxiv.org/abs/2607.01967}
}

@misc{vazquez2026communitystructurepseudofractalweb,
  title = {Community structure of the pseudofractal web},
  author = {Vazquez, Alexei},
  year = {2026},
  eprint = {2607.03010},
  archivePrefix = {arXiv},
  primaryClass = {physics.soc-ph},
  url = {https://arxiv.org/abs/2607.03010}
}

@article{reichardt2004,
  author = {Reichardt, J\"org and Bornholdt, Stefan},
  title = {Detecting Fuzzy Community Structures in Complex Networks with a {Potts} Model},
  journal = {Phys. Rev. Lett.},
  volume = {93},
  pages = {218701},
  year = {2004},
  publisher = {American Physical Society},
  doi = {10.1103/PhysRevLett.93.218701},
  url = {https://doi.org/10.1103/PhysRevLett.93.218701}
}

\end{document}